# Minimum relaxation n-body simulations using orthogonal series force computation

Martin D. Weinberg[1]

Department of Physics and Astronomy

University of Massachusetts, Amherst, MA 01003-4525

weinberg@phast.umass.edu

## ABSTRACT

This report describes a modification of the orthogonal function Poisson solver for n-body simulations that minimizes relaxation caused by small particle number fluctuations. With the standard algorithm, the noise leading to relaxation can be reduced by making the expansion basis similar to the particle distribution and by carefully choosing the maximum order in the expansion. The proposed algorithm accomplishes both tasks simultaneously while the simulation is running. This procedure is asymptotically equivalent to expanding in an orthogonal series which is matched to the distribution to start and truncating at low order. Because the modified algorithm adapts to a time-evolving distribution, it has advantage over a fixed basis.

The required changes to the standard algorithm are minor and do not affect its overall structure or scalability. Tests show that the overhead in CPU time is small in practical applications. The decrease in relaxation rate is demonstrated for both axisymmetric and non-axisymmetric systems and the robustness of the algorithm is demonstrated by following the evolution of unstable generalized polytropes. Finally, the empirically based moment analysis which leads to the uncorrelated basis is an ideal tool for investigating structure and modes in n-body simulations and an example is provided.

## 1. Introduction

No n-body stellar dynamical system has zero relaxation; relaxation results from Poisson fluctuations in the coarse-grained mass density and is an intrinsic feature of the underlying physics. Because n-body galaxy simulations have orders of magnitude fewer bodies than in reality, the relaxation rates are artificially high. A variety of techniques have been devised to help ameliorate this problem, and most strategies trade-off spatial resolution for decreased relaxation.

---

[1]Alfred P. Sloan Foundation Fellow.

For example, to understand the dynamical evolution of galaxies on the largest scales, one may fit the density and potential by functions which only vary on the scales of interest and thereby suppress this unwanted fluctuation. If the distribution is well-represented by a small number of terms, the overall power in fluctuations, and therefore the relaxation, will be smaller. This motivation is behind a number of approaches described by Clutton-Brock (1972, 1973), Kalnajs (1976), Fridman & Polyachenko (1984), Allen et al. (1990), and most recently by Hernquist & Ostriker (1992, HO) all of whom derive orthogonal series—potential–density pairs more precisely—to represent the mass distribution. We follow HO in calling this a *self-consistent field* (SCF) algorithm. Earn & Sellwood (1995) discuss the virtues of this approach for studying global modes.

The few lowest order members of the ideal basis fully represent the underlying distribution. Unfortunately, it is impractical to analytically construct specialized bases for most applications. Because a general profile is unlikely to be a member of the basis, the quality of the force computation will depend on the series truncation. If one truncates too early, the density will be poorly represented (bias) and the resulting evolution will not be accurately obtained. If truncates too late, high-order fluctuations will increases the noise (variance) and lead to large relaxation.

Accurate representation of the gravitational field based on a sample of discrete points is an example of the more general statistical problem of density estimation and SCF is a special case of an *orthogonal series density estimator*. Tarter and Kronmal (1970, hereafter TK) derive several theorems for density estimation by Fourier expansion and suggest series termination or stopping criteria based on the mean integrated squared error (MISE). The essence of this technique is intuitively clear: the estimated values of series coefficients are compared to their variances and only included in the sum when the signal-to-noise ratio is larger than unity. These criteria have been further studied and improved by Hall (1981, see Izenman 1991 for a review). The improved approach extremizes the MISE to find optimal trade-off between errors in bias and variance. The end result is a set of weights that are unity for high signal-to-noise and zero for low signal-to-noise coefficients. These results are straightforwardly generalized to complete but non-Fourier basis sets. Merritt & Tremblay (1994) recently discussed another rigorous approach to estimating a smooth density and potential from a particle distribution.

The algorithm derived here further improves the series estimator by empirically determining a best-fitting, uncorrelated basis before applying Hall's method. This prevents loss of globally correlated signal which finds itself distributed among multiple components with low signal-to-noise due to a poor choice of basis. In addition, the analysis determines the statistical significance of apparent features in an n-body simulation, which was the original motivation behind this study. Application to SCF will be derived below (§2) followed by test applications (§3). The algorithm is summarized in the Appendix together with some implementation detail.

## 2. A modified orthogonal series potential solver

An orthogonal series density estimator in one dimension is defined through its coefficients as follows:

$$a_j = \int dx \, \bar{\Psi}_j(x) f(x) \tag{1}$$

where

$$\int dx \, \bar{\Psi}_j(x) \Psi_k(x) = \delta_{jk} \tag{2}$$

and the distribution $f(x)$ is well-defined but unknown a priori. For a particular sample of $N$ points, the coefficients based on equation (1) are

$$\hat{a}_j = \frac{1}{n} \sum_{k=1}^{n} \Psi_j(x_k). \tag{3}$$

Finally, taking $\hat{a}_j$ as estimates of $a_j$, the estimate for the distribution $f(x)$ becomes

$$\hat{f}(x) = \sum_{j=1}^{M} \hat{a}_j \Psi_j(x). \tag{4}$$

This procedure is straightforwardly generalized to 2 or 3 dimensions and to biorthogonal functions. In particular, for a spherical SCF expansion, $\Psi_{lm}^j = p_{lm}^j(r) Y_{lm}(\theta, \phi)$ and $\bar{\Psi}_{lm}^j = d_{lm}^j(r) Y_{lm}(\theta, \phi)$ where $d_{lm}^j/4\pi G$ and $p_{lm}^j$ are the biorthogonal potential-density pairs. Potential-density pairs solve the Poisson equation individually and therefore eliminate the otherwise necessary step of determining the potential or force from the density. These pairs are well-described in the literature (cf. §1). For simplicity of notation, we will use one-dimensional examples.

### 2.1. Optimal convergence

In order to choose the appropriate number of coefficients in equation (4), we need an appropriate measure to quantify the goodness of the estimate for $f(x)$. Some of the earliest investigators used the mean integrated square error (MISE) defined as follows:

$$D \equiv \int dx \, E\left[\hat{f}(x) - f(x)\right]^2 \tag{5}$$

where $E\left[\cdot\right]$ denotes expectation value. As described in §1, too few higher-order terms lead to a poor approximation $\hat{f}$ and to a larger than optimal value of $D$. Similarly, too many high-order terms lead to a noisy approximation and also to a larger than optimal value of $D$. This led Tarter & Kronmal (1970) to derive an expression for the growth of the MISE as a single term is added. If the MISE decreases, the term is kept in the sum and vice versa. This can be generalized by an expansion of the following form:

$$\hat{f}(x) = \sum_{j=1}^{M} b_j \hat{a}_j \Psi_j(x) \tag{6}$$

where the $b_j$ are *smoothing coefficients* and $M$ is some practically determined upper limit. The MISE for this form is easily shown to be

$$D(\mathbf{b}) = \sum_j \left[b_j^2 \mathrm{var}(\hat{a}_j) + (b_j - 1)^2 a_j^2\right] \tag{7}$$

(Hall 1986) where $\mathrm{var}(\hat{a}_j)$ is the variance of the estimator $\hat{a}_j$. Term by term, the MISE is minimized for

$$b_j = \left[1 + \mathrm{var}(\hat{a}_j)/a_j^2\right]^{-1}. \tag{8}$$

Hall (1986) derives bounds for the convergence of $D(\mathbf{b})$ for this optimal choice and several related choices for $b_j$. The computation of $\mathrm{var}(\hat{a}_j)$ is discussed in the Appendix. Equations (6) and (8) motivate choosing $M$ large enough so that $b_M$ remains very small throughout the simulation. If we interpret $S/N \equiv [\hat{a}_j^2/\mathrm{var}(\hat{a}_j)]^{1/2}$ as the signal-to-noise ratio, then the marginally significant feature with $S/N = 1$ has $b_j = 0.5$. Experience suggests that features with $S/N \approx 0.7, b \approx 0.3$ are also discriminated.

## 2.2. Transformation to principal components

We could now proceed with the force computation using the estimator and smoothing algorithm from the previous section. However even in the absence of dynamics, the fluctuations in the components $\hat{a}_j$ will be correlated since a particular fluctuation will not be represented by an arbitrary basis member $\Psi_j$. In other words, the correlation matrix for the coefficients of an arbitrary orthogonal basis will not be diagonal. Smoothing the coefficients in the fiducial basis, then, will help reduce the MISE but will only succeed in removing *parts* of any fluctuation due to correlated noise. Conversely, a particular component in a statistically dependent basis may have a low signal-to-noise ratio itself but support a statistically significant signal on the whole.

For these reasons, we would like a basis whose members best represent the distribution with the fewest number of terms, Condition (i), and which decorrelates the variation of individual components, Condition (ii). Such a basis may be directly obtained using the particle distribution as follows. Let $\rho_i = \{\Psi_1(x_i), \Psi_2(x_i), \ldots, \Psi_M(x_i)\}$, the weight of an individual particle to the estimate of the coefficient vector $\hat{\mathbf{a}}$. Condition (i) suggests the best representation in the space spanned by the $M$ basis functions are the vectors $\mathbf{e}$ which maximize

$$\chi^2 = \frac{1}{N} \sum_{i=1}^{N} (\mathbf{e} \cdot \rho_i)^2 \tag{9}$$

subject to a normalization condition which we will take to be $\mathbf{e}^T \cdot \mathbf{e} = 1$. This extremization problem can be solved using Lagrange multipliers with the solution

$$\mathbf{S} \cdot \mathbf{e} = \lambda \mathbf{e} \tag{10}$$

where $\mathbf{S}$ is the matrix

$$S_{kl} = \frac{1}{N} \sum_{i=1}^{N} \Psi_k(x_i) \Psi_l(x_i) \tag{11}$$

and $\lambda$ is the Lagrange multiplier. Because equation (10) defines an eigenvalue problem and $\mathbf{S}$ can be written as a real, symmetric matrix, this is readily solvable with real eigenvalues $\lambda$. Therefore, equation (9) has $M$ real orthonormal solutions $\mathbf{e}^{[k]}$. Let $T_{jk}$ be the matrix whose columns are $\mathbf{e}^{[k]}$. This matrix is unitary and defines a basis transformation.

A geometric interpretation of this transformation helps make its meaning clear. Each eigenvector defines a mutually orthogonal direction in the $M$ dimensional coefficient space. Let us order the solutions to equation (10) by eigenvalue from largest to smallest. The sum of the projections squared for all $\rho_i$ on to the first eigenvector is the maximum of any possible direction and therefore best represents the particle distribution in the root mean squared sense (cf. eq. 9). Next, the sum of the projections squared on to the second eigenvector is the maximum of any possible direction but the one defined by the first eigenvector. Continuing, the sum of the projections squared on to the third eigenvector is the maximum of any possible direction but the subspace spanned by the first and second eigenvector, and so forth until the space is fully spanned. The transformation $\mathbf{T}$, then, is the optimal decomposition which best accounts for the particle distribution with the smallest number of terms, satisfying Condition (i).

The cross-product matrix $\mathbf{S}$ describes the correlation between members of the original basis. A similarity transformation by $\mathbf{T}$ diagonalizes $\mathbf{S}$ and therefore describes a transformation from the original basis to a new basis in which the members are uncorrelated, satisfying Condition (ii). Finally, the sample mean of the $\rho_i$ is equivalent to the density estimation problem defined in equations (3) and (4), $\hat{\mathbf{a}} = 1/N \sum_{i=1}^{N} \rho_i$ and the matrix projects this to the new uncorrelated best-fitting basis.

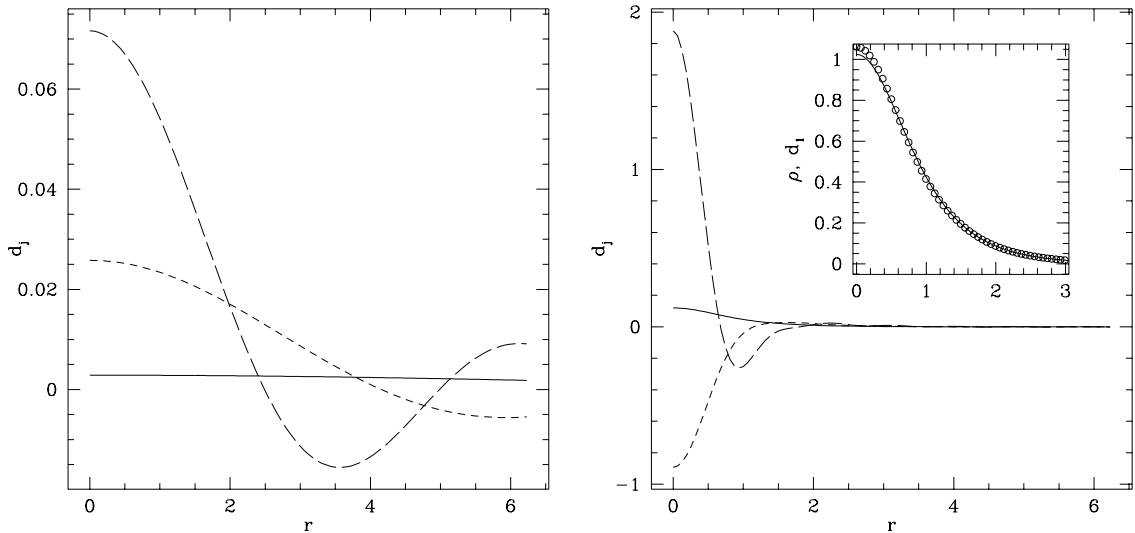

Fig. 1.— Three lowest order density members of the original expansion basis (left) compared with the derived set (right). Members $j = 1, 2, 3$ are shown with solid, short-dashed and long-dashed lines, respectively, The inset demonstrates that the lowest order term in the derived set (open circles) closely approximates the background profile itself (solid line).

A search revealed that this approach is called *empirical orthogonal function* analysis in the meteorological literature and appears to date back to Lorenz (1956, see Glahn 1985 for a review). In weather prediction, one often has a grid of data from ground stations and wants to determine any patterns of variation over nominal values in temperature, pressure, etc. with time. Lorenz showed that the members of the transformed basis account for the largest fraction of total original variance. Analogously, we may define the relative goodness of the representation truncated after $k < M$ terms as

$$F_k = \frac{\sum_{j=1}^{k} \lambda_j}{\sum_{j=1}^{M} \lambda_j} \equiv \sum_{j=1}^{k} \lambda_j'. \tag{12}$$

The better the particle distribution is fit, the faster $F_k$ approaches unity as $k$ increases.

Applied to the problem at hand, the lowest order basis function will look like the underlying particle distribution, the next function will represent the largest scale fluctuation about the background and so on. Take as an example an isotropic low-concentration King model ($W_0 = 3$), Monte Carlo realized by 20000 mass points. The first panel in Figure 1 shows the lowest order components of the fiducial spherical Bessel function basis and the second panel shows the transformed basis. The lowest order transformed function (solid) looks like the underlying profile as expected (see inset) and carries 99.9% of the total weight. We may now apply the optimal smoothing formula in the new basis to derive the coefficients $b_j$. The cumulative weight $F_k$ and the smoothing coefficients $b_k$ are shown in Figure 2. Notice that the $b_j$ are nearly zero for all high-order components which therefore make no contribution.

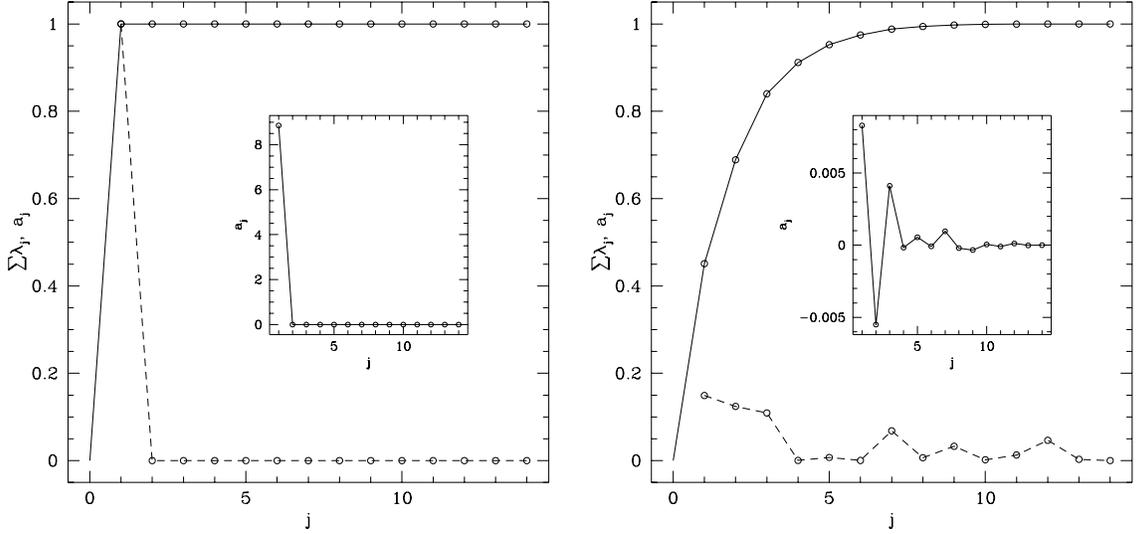

Fig. 2.— Solid curve: cumulative weight for new basis, $\sum_{i=1}^{k} \lambda'_j$. Dashed curve: optimal smoothing $b_j$ for new basis coefficients. Inset: values of unsmoothed new basis coefficients. Plot on left (right) is for the $l = m = 0$ ($l = m = 2$) harmonic. The low values for both $\hat{a}_j$ and $b_j$ for $l = m = 2$ indicate that this harmonic does not contribute, as expected.

HO point out that fluctuations will be suppressed and relaxation minimized if the lowest order basis function is the underlying density function. In principle, the combined procedure advanced here—basis transformation followed by optimal smoothing—achieves the same effect. Furthermore, the procedure is inherently adaptive and so will continue to work even after significant evolution removes the advantage of an ab initio tuned but fixed basis. The only constraint on the underlying orthogonal functions is that they solve the appropriate boundary conditions for the physical system. In practice, this is a weak constraint. In addition, analysis of the transformed independent basis functions provide an empirical approach to understanding the underlying modes and is likely to further improve the approach advanced by Earn & Sellwood (1995).

I advocate applying the algorithm separately for each harmonic order. One can extend the procedure to all harmonic orders simultaneously and obatain a good representation. However, tests show that this fit is more sensitive to smaller-scale local features and is therefore less efficient at suppressing fluctuation-driven relaxation and less useful for discerning global modes and distortions. In addition, because the correlation matrix **S** is full rather than block diagonal, the extended procedure is more computationally intensive.

## 3. Numerical tests

The procedure outlined in §2 is designed to best extract all information in the particle distribution. This means that some of the structure due to unrealistically small numbers of particles is real will be retained. Weinberg (1993) discusses these intrinsic modes which lead to relaxation. Some over smoothing may reduce the low-amplitude disturbances which cause the relaxation. Here we propose a simple modification to optimal smoothing:

$$b_j = [1 + \alpha \operatorname{var}(\hat{a}_j)/a_j^2]^{-1} \tag{13}$$

for some $\alpha \geq 1$. The quantity $a_j^2/\operatorname{var}(\hat{a}_j)$ is a measure of the signal-to-noise ratio squared and therefore increasing $\alpha$ will over smooth the profile and reduce low-amplitude fluctuations. We will explore the practical application of optimal smoothing and over smoothing below.

### 3.1. Equilibrium models

#### 3.1.1. Spherical

HO presented several relaxation measures and here we will adopt the r.m.s. change in relative orbital energy per particle, $\langle(\Delta E/E)^2\rangle$. Figure 3 shows this measure for a simulation of an isotropic $W_0 = 5$ King model with $N = 20000$ for 3 different cases of smoothing. The maximum radial order of the Bessel function basis is $n_{max} = 14$ and maximal harmonic order is $l_{max} = 4$. The standard algorithm with no smoothing (dotted) is compared with optimal smoothing (solid) discussed in the previous section and over smoothing (dashed) with $\alpha = 16$. Note the $\langle(\Delta E/E)^2\rangle \propto t$ as expected for relaxation. The relative relaxation rate is therefore $d\langle(\Delta E/E)^2\rangle/dt$ and is estimated from the slopes in Figure 3. The optimally smoothed relaxation rate is lower than the unsmoothed case but larger than the over-smoothed case. The relaxation rate in the over-smoothed case is nearly a factor of 3 smaller than the unsmoothed case and largely results from the amplitude shifts in the central value of the potential; this is physical and impossible to eliminate. The rates for a fixed scheme with increasing values of particle number, $N$, scale slightly faster ($N^{-1.25}$) than the naively expected $N^{-1}$; the non-Poisson behavior is due to the dependence of the smoothing algorithm on $N$.

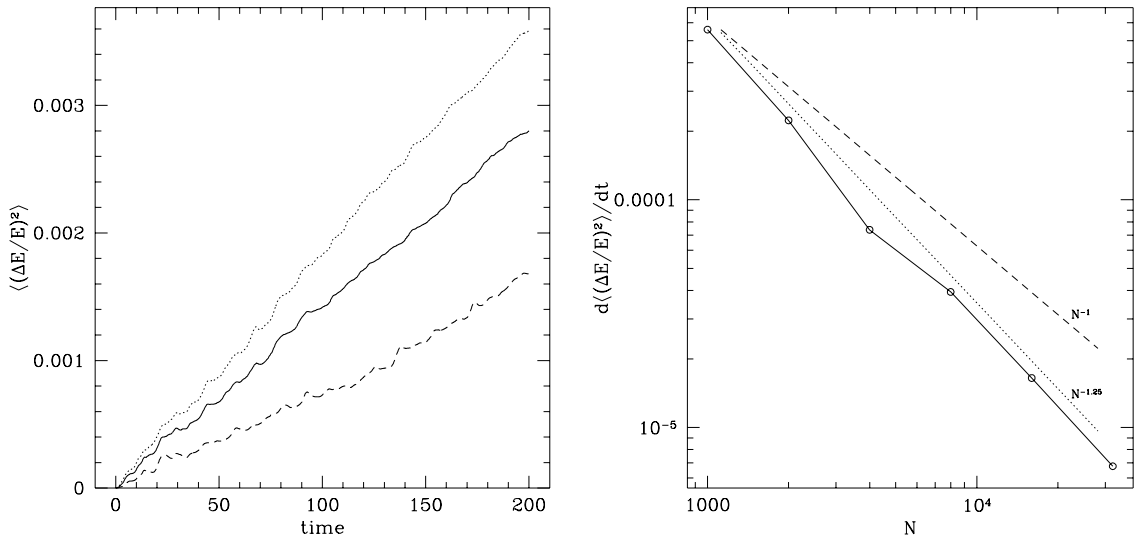

Fig. 3.— Change in relative energy per particle for the $W_0 = 5$ King model ($N = 20000$) without the proposed algorithm (dotted), with optimal smoothing (solid), and with $\alpha = 16$ (dashed).

Fig. 4.— Scaling of relaxation rate with particle number $N$ (open circles). Best fit power law $N^{-1.25}$ (dotted) and $N^{-1}$ (dashed) shown for comparison.

Figure 5 shows similar tests for a Plummer sphere using both Clutton-Brock and Hernquist bases which match the outer boundary conditions for the infinite extent, finite mass Plummer model. The relaxation rates group separately for the optimal and over-smoothed algorithm (listed in Table 1) and smoothing decreases the rate in all cases. The relative improvement from the modified algorithm is much larger than for the King model (cf. Fig. 3). The lowest rate obtains for $l_{max} = 0$, optimally smoothed, but is similar in rate to the over-smoothed cases. The over-smoothed algorithm with $l_{max} = 4$ results in a relaxation rate that is a factor of 5 smaller than the direct Clutton-Brock expansion.

For $l_{max} = 4$, the unsmoothed algorithm results in nearly identical relaxation with both bases, consistent with relaxation dominated by non-axisymmetric disturbances. The lowest order basis function in the Clutton-Brock set is coincident with the unperturbed Plummer model. Note that both the smoothed Hernquist set and Clutton-Brock set with give very similar results, suggested that the modified algorithm recovers full benefit of the tailored basis set.

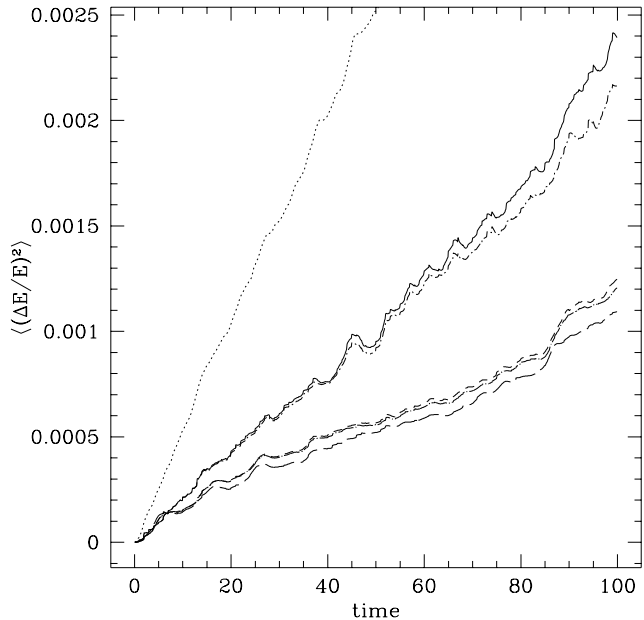

Fig. 5.— Relaxation for a Plummer model ($N = 20000$) without the proposed algorithm using the Clutton-Brock basis (dotted), using optimal smoothing (solid), using over smoothing (short dash), optimal smoothing with $l_{max} = 0$ (long dash), Hernquist basis with optimal smoothing (dot–short dash), Hernquist basis with over smoothing (dot–long dash).

Because fluctuation-driven relaxation is part of the true solution to the n-body problem, over smoothing ($\alpha > 1$) may modify the underlying dynamics. Although larger than unity values of $\alpha$ improves the long-term coherence of individual orbits, the over smoothing could artificially suppress, for example, a weakly unstable disturbance and therefore should be used with caution. In the cases in Table 1, optimal smoothing reduces the relaxation rate by a factor of three and is the conservative choice. In the test cases discussed here, the relaxation rate decreases $\alpha$ is increased.

Table 1: Relaxation rates for Plummer models ($N = 20000$)

| Basis | $n_{max}, l_{max}$ | Smoothing | Rate |
|---|---|---|---|
| Clutton-Brock | 14, 0 | none | $1.1\,e-05$ |
| Clutton-Brock | 14, 0 | optimal | $8.0\,e-06$ |
| Clutton-Brock | 14, 4 | none | $5.1\,e-05$ |
| Clutton-Brock | 14, 4 | optimal | $2.1\,e-05$ |
| Clutton-Brock | 14, 4 | $\alpha = 16$ | $8.6\,e-06$ |
| Hernquist | 14, 0 | none | $1.1\,e-05$ |
| Hernquist | 14, 0 | optimal | $7.8\,e-06$ |
| Hernquist | 14, 4 | none | $4.8\,e-05$ |
| Hernquist | 14, 4 | optimal | $1.9\,e-05$ |
| Hernquist | 14, 4 | $\alpha = 2$ | $1.7\,e-05$ |
| Hernquist | 14, 4 | $\alpha = 4$ | $1.3\,e-05$ |
| Hernquist | 14, 4 | $\alpha = 8$ | $1.1\,e-05$ |
| Hernquist | 14, 4 | $\alpha = 16$ | $8.3\,e-06$ |

### 3.1.2. Axisymmetric disk

Simulations of exponential disks embedded in rigid equal mass King model $W_0 = 3$ halos for stability, verified that a similar decrease in relaxation rate also obtains in two-dimensions. The magnitude of the decrease using optimal smoothing was similar although slightly smaller, factors of 2—3 rather than 3—4 found in three-dimensions. Unlike the spherical cases, over smoothing made little further improvement. Because of the proposed algorithm adaptively constructs a new orthogononal functions, a fiducial Bessel function basis should suffice for finite extent two-dimensional disk models. This should be easily extendible to a three-dimensional thick disk geometry, although this has not been tested here.

The ease of graphically representing the two-dimensional reconstruction provides a demonstration of the proposed algorithm. Figure 6 shows the isodensity contours of an exponential disk of 80000 particles using Clutton-Brock's biorthogonal functions with maximum radial and angular orders $n_{max} = 10$ and $m_{max} = 4$. The non-axisymmetric noise is clearly seen using the standard algorithm and suppressed by more than a factor of 100 using the new algorithm.

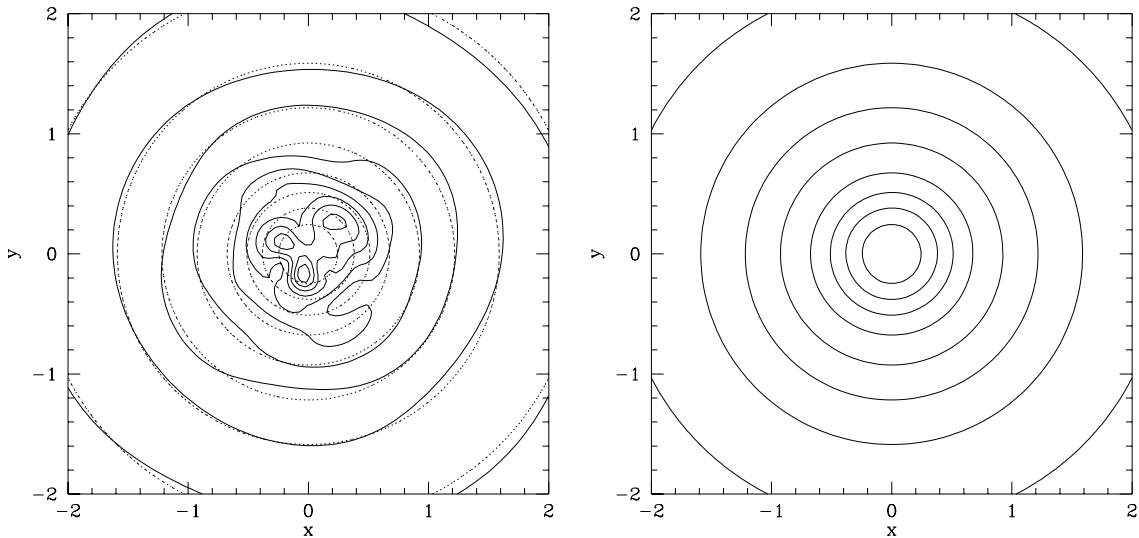

Fig. 6.— Isodensity contours for an $N = 80000$ particle Monte Carlo realization of an exponential disk. Left: Density reconstruction using standard algorithm (solid) compared with smoothed algoritm (dotted). Right: reconstruction using smoothed algorithm alone for comparison.

### 3.1.3. Non-spherical

The same diagnostics were applied to prolate systems to make sure that the behavior in the previous section holds for non-axisymmetric cases. Distribution functions for the perfect prolate models (de Zeeuw 1985) with $a = 1, b = c = 0.2$ were generated with atlas technique following Statler (1987).

Figure 7 shows the increase in $\langle (\Delta E/E)^2 \rangle$ with time. The initial period of growth ($t \lesssim 10$) is caused by inaccuracy in the equilibrium solution due to a finite atlas; the mean squared change in energy is approximately 8% during the first few dynamics times as the model "mixes" to a self-consistent equilibrium. Thereafter, one sees slow linear growth due to relaxation. Growth rates are derived from fits to the curves for $t > 10$ in Figure 7 and shown in Table 2. The new algorithm relaxation rate decreases by a factor of 3.5 to 3.9.

The last four cases in Table 2 illustrate the effects of over smoothing. The relative relaxation rate decreases as $\alpha$ increases but the increase in $\langle (\Delta E/E)^2 \rangle$ during the initial "mixing" very slowly increases with increasing $\alpha$ due to elimination of some significant non-axisymmetric structure. The relative decrease from $\alpha = 1$ to $\alpha = 16$ is nearly 40%. All in all, the performance of the modified algorithm for the non-axisymmetric prolate model is similar to that for the spherical models.

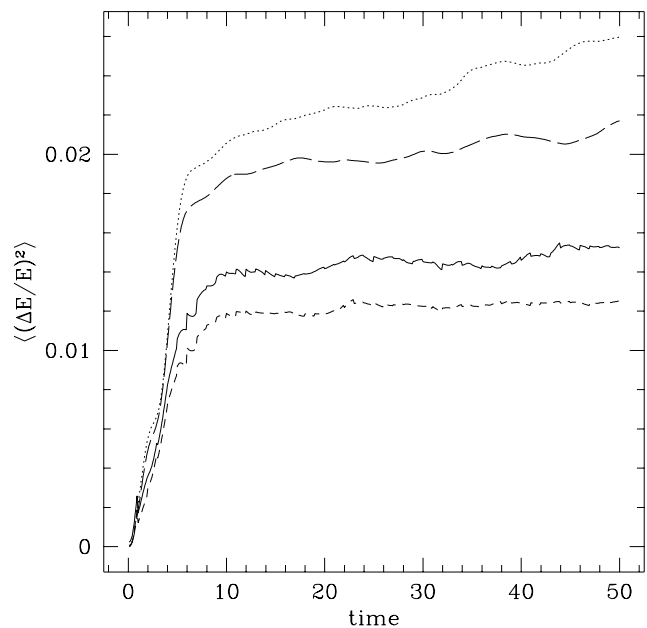

Fig. 7.— Relaxation for a perfect prolate models shown for $N = 20000$ with and without smoothing (solid and dotted) and for $N = 80000$ (short dash and long dash).

Table 2: Relaxation rates for perfect prolate models

| Basis | $n_{max}, l_{max}$ | Particles | Smoothing | Rate |
|-------|---------|-----------|-----------|------|
| Clutton-Brock | 14, 8 | 20000 | none | $1.3\,e - 04$ |
| Clutton-Brock | 14, 8 | 20000 | optimal | $3.3\,e - 05$ |
| Clutton-Brock | 14, 8 | 80000 | none | $5.5\,e - 05$ |
| Clutton-Brock | 14, 8 | 80000 | optimal | $1.6\,e - 05$ |
| Clutton-Brock | 14, 8 | 80000 | $\alpha = 2$ | $1.6\,e - 05$ |
| Clutton-Brock | 14, 8 | 80000 | $\alpha = 4$ | $1.4\,e - 05$ |
| Clutton-Brock | 14, 8 | 80000 | $\alpha = 8$ | $1.2\,e - 05$ |
| Clutton-Brock | 14, 8 | 80000 | $\alpha = 16$ | $1.0\,e - 05$ |

## 3.2. Unstable models

Because the optimal smoothing procedure diminishes the amplitude of low signal-to-noise features, one needs confidence that otherwise growing modes will not be suppressed. For tests, we chose the generalized polytropes investigated by Barnes, Goodman, & Hut (1986) and examined the evolution of a sequence with fixed $N = 0.75$ and $-0.8 < M < 0$. The $M = 0$ model is isotropic and the model is more radially anisotropic with decreasing $M$. For $M \leq -0.25$, the velocity distribution becomes bimodal which is the stability boundary proposed by Barnes et al..

The instability develops rapidly in the most unstable model tested, $M = -0.8$, for both the smoothed and unsmoothed algorithm with $N = 20000$. The growing modes and final states are in good agreement in both cases, although the structure develops in different directions because the noise spectrum is different. As $M \to -0.25$, the growth of the mode is delayed in the smoothed relative to the unsmoothed algorithm due to the lower-amplitude noise spectrum. For $M = -0.3$, there is no growing instability evident for after 40 half-mass crossing times although a time-series analysis of the $m = 2$ coefficients suggests that the mode is appearing and disappearing transiently. This might be expected very close to the boundary.

A simulation with $N = 0.75, M = -0.4$ using the over-smoothed ($\alpha = 12$) algorithm develops the $m = 2$ instability within 25 crossing times which the same as the measured rate for the optimally smoothed ($\alpha = 1$) case. This suggests that moderate over smoothing does not adversely affecting the modal structure, but that weak modes most likely *will* be modified.

Finally, it is difficult to assess the importance of a low-level feature in a simulation. As a by-product of optimal smoothing, the principal component transformation accurately picks out the mode, and the smoothing coefficients $b_j$ for the transformed basis indicate the statistical significance of the mode. As an example, Figure 8 shows a well-evolved instability for $N = 0.75, M = -0.5$ in the principal axis plane (left) together with the non-axisymmetric component only (right). Only the lowest-order radial functions for $l = m = 0$ and $l = m = 2$ angular harmonics are significant according to the smoothing algorithm. Figure 9 shows the mode at an earlier time; one might miss the disturbance in a visual inspection but the analysis reveals that lowest-order $l = m = 2$ component is significant. The non-axisymmetric profile in Figure 9 is comparable to the saturated mode a later times, Figure 8.

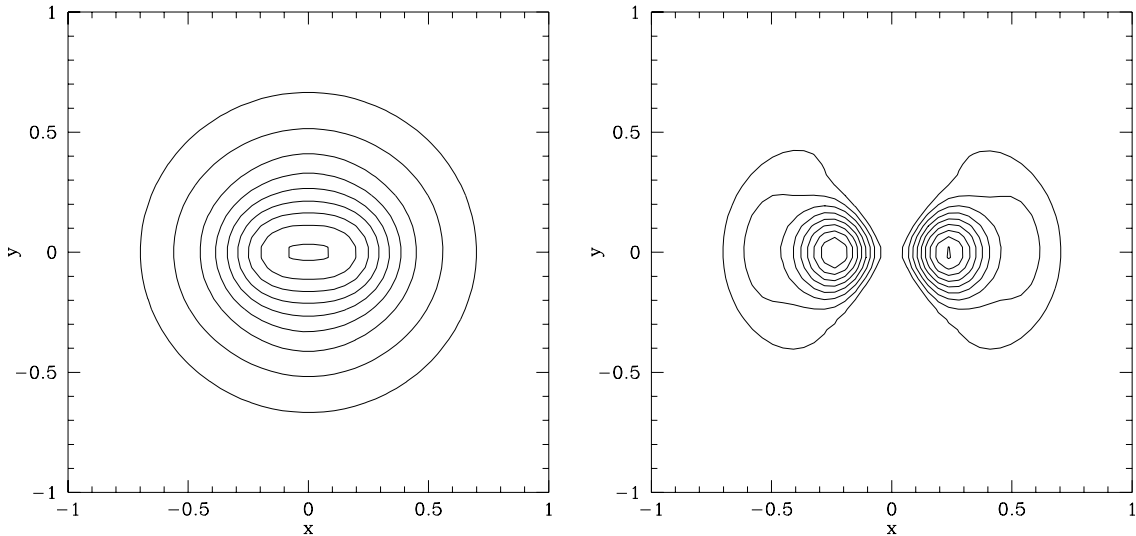

Fig. 8.— Left: Shape of growing instability for $N = 0.75, M = -0.5$ generalized polytrope in the principal axis coordinate system. The shape is dominated by two basis functions in the transformed basis: the lowest-order radial functions $l = m = 0$ and $l = m = 2$. Right: the non-axisymmetric contribution to the left-hand panel.

## 4. Summary

Orthogonal series force computation (the SCF method) reduces relaxation by limiting the range of spatial variation to large scales. Techniques from statistical density estimation are used to derive an optimal smoothing algorithm for SCF which in essence selects the minimum statistically significant length scale. This is combined with empirically determined orthogonal functions that best represent the particle distribution and separate any correlated global patterns. The overall approach may have application to the more general problem of orthogonal series density estimation.

The procedure provides nearly all of the benefit of an analytically derived basis set which matches the underlying profile at lowest order with the advantage of adaptively following the subsequent evolution. We found that optimal smoothing decreases the relaxation rate by at least a factor of three in most cases. Further smoothing reduces relaxation rate even farther in three-dimensional simulations with only minor loss of resolution.

The algorithm adaptively constructs a statistically uncorrelated orthogonal basis from a fiducial orthogonal basis. The fiducial basis is only required to span enough of the function space to be capable of representing the particle distribution. The lowest order member in the new basis represents the underlying profile and successive members represent the dominant disturbance. As a by product, therefore, the new basis is a diagnostic tool for the evolution of structure in simulations and will be useful for identifying multiple growing and damped modes and characterizing forced responses.

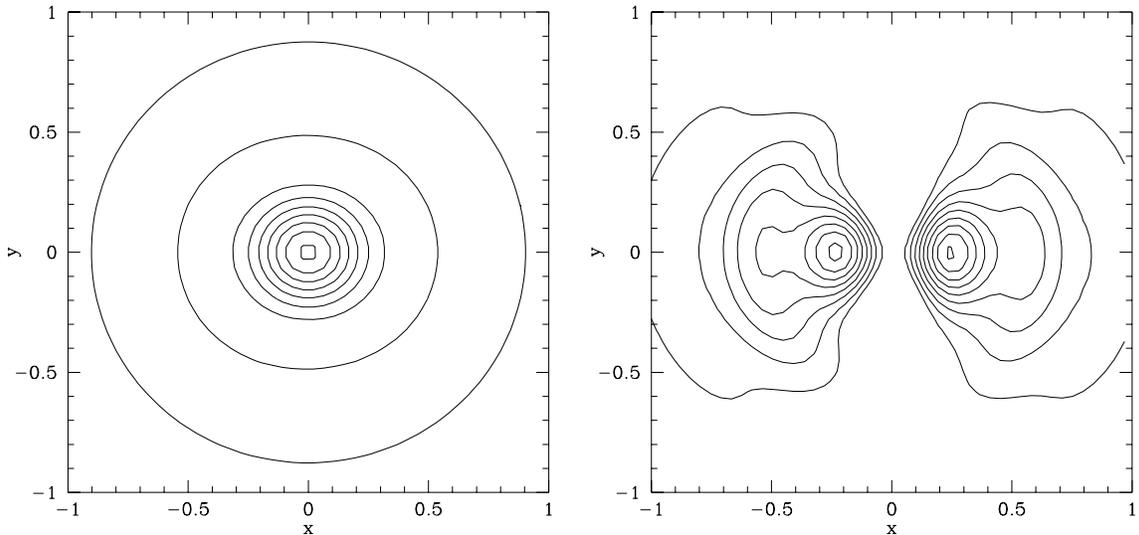

Fig. 9.— As in Fig. 8 but at an earlier time. This $l = m = 2$ feature has $S/N \approx 1$.

I thank Sergai Nikolaev for discussions. This work was supported in part by NASA grant NAGW-2224, NAG 5-2873 and the Sloan Foundation.

## A. The modified algorithm: summary statement

1. Accumulate contribution to coefficients $\hat{a}_j$ and correlation matrix $S_{jk}$ in a loop over all particles (eqs. 3 and 11). This step is common to the original SCF algorithm.

2. Solve the eigenvalue problem (eq. 10) in order to construct the transformation $\mathbf{T}$ to the best-fitting uncorrelated basis. If $\mathbf{e}^{[k]}$ are the eigenvectors with eigenvalues $\lambda_j$, $T_{jk} = e_j^{[k]}$ defines the transformation.

3. Transform the coefficient estimates $\hat{a}_j$ to the new basis (denoted by star), $\hat{a}_j^\star = \sum_k T_{jk}\hat{a}_k$, and use $\mathrm{var}(\hat{a}_j^\star) = \lambda_j - \hat{a}_j^{\star 2}$ to compute $b_j^\star$ following equation (8).

4. The modified estimate $\hat{f}^\star(x)$ then follows from equation (6) where non-starred are replaced by starred quantities with $\Psi_j^\star(x) = \sum_k T_{jk}\Psi_k(x)$.

## B. Variance computation

The variance of the estimates $\hat{a}_j$ follows from the definitions in §2 and is

$$
\begin{aligned}
\mathrm{var}(\hat{a}_j) &\equiv E\left[\hat{a}_j - a_j\right]^2 \\
&= E\left[\frac{1}{N^2}\sum_{s,t=1}^N \Psi_j(x_s)\Psi_j(x_t) - \frac{1}{N}\sum_{s=1}^N \Psi_j(x_s)\frac{1}{N}\sum_{t=1}^N \Psi_j(x_t)\right]
\end{aligned}
\tag{B1}
$$

having used $a_j = E[\hat{a}_j]$ to obtain the second equality. The first term in this expression may be simplified by splitting the double sum over particle index:

$$
\begin{aligned}
E\left[\frac{1}{N^2}\sum_{s,t=1}^N \Psi_j(x_s)\Psi_k(x_t)\right] &= E\left[\frac{1}{N^2}\sum_{s=1}^N \Psi_j(x_s)\Psi_k(x_s) + \frac{1}{N^2}\sum_{t\neq s}^N \Psi_j(x_s)\Psi_k(x_t)\right] \\
&= \frac{1}{N}E\left[\frac{1}{N}\sum_{s=1}^N \Psi_j(x_s)\Psi_k(x_s)\right. \\
&\qquad \left. + (N-1)\frac{1}{N}\sum_{s=1}^N \Psi_j(x_s)\frac{1}{N}\sum_{t=1}^N \Psi_k(x_t)\right] \\
&= \frac{1}{N}\left\{E\left[\Psi_j\Psi_k\right] + (N-1)E[\Psi_j]E[\Psi_k]\right\}
\end{aligned}
\tag{B2}
$$

where the mass of particle of each particle is taken to be $1/N$ and $E[\cdot]$ is denotes the expectation value. The sample value of the first term in the expression above is the matrix $S_{ij}$ defined in §2.2. Combining equation (B2) with equation (B1) yields the variance in a familiar form:

$$
\mathrm{var}(\hat{a}_j) = \frac{1}{N}\left\{E\left[\Psi_j\Psi_j\right] - E[\Psi_j]E[\Psi_j]\right\}.
\tag{B3}
$$

The expected quantities are estimated by the sample quantities in the smoothing procedure. The quantities $E[\Psi_j]$ are the coefficients $a_j$ and $E[\Psi_j\Psi_k]$ is the second moment matrix $S_{jk}$. The computation of $E[\Psi_j\Psi_k]$ is trivially vectorizable and parallelizable following the same procedure used for the coefficients.

## C. Computational detail

The computation of eigenvectors $T_{jk}$ will have very small overhead for most applications (and was performed in the tests described here by Householder's method and the QL algorithm). All of the overhead in the algorithm is in the computation of the moment matrix in the first term of equations (B2) and (B3). If the variance is computed at every step, the overhead will be order unity. There are several obvious remedies: 1) one may make use of the asymptotic properties of one's basis to estimate the variance without direct computation; 2) because each particle's contribution to the covariance is essentially an outer vector product, one may use special-purpose vector hardware to good advantage; and 3) one may compute the variance after multiple time steps since the dominant low-order correlations are have a longer time scale than a single time step. We adopted the latter prescription here which requires saving the coefficients $b_j$ and transformation $T_{jk}$ for use in intermediate time steps. Tests with $b_j$ and $T_{jk}$ recomputation every time step (approximately one hundredth of a characteristic crossing time) and every ten time steps gave nearly identical relaxation rates. Even larger intervals may be possible depending on the rate of evolution.

Because the proposed algorithm is adaptive and minimizes the MISE, a larger $n_{max}$ may yield a better estimate (lower MISE) and without risk of adding noise as in the standard SCF algorithm. The trade-offs in performance will be both application and implementation dependent. For the tests in §3, doubling $n_{max}$ from 14 to 28 had negligible effect on relaxation rates.